\documentclass[twocolumn,showpacs,superscriptaddress,aps,pra]{revtex4}
\usepackage{epsfig}
\usepackage{subfigure}
\usepackage{amssymb}
\usepackage{graphicx}
\usepackage{color}
\usepackage{amsfonts}
\usepackage{amscd}
\usepackage{hyperref}
\usepackage{amsmath}
\usepackage{epstopdf}

\begin{document}

\newcommand{\cmt}[1]{{\textcolor{red}{[#1]}}}
\newcommand{\qn}[1]{{\textcolor{red}{ (?)  #1 }}}
\newcommand{\chk}[1]{{\textcolor{green}{#1}}}
\newcommand{\del}[1]{{\textcolor{blue}{ \sout{#1}}}}
\newcommand{\rvs}[1]{{\textcolor{blue}{#1}}}%crt means "revision"

\title{Cross-Kerr effect on an optomechanical system}

\author{Wei Xiong}

\affiliation{Department of Physics, Fudan University, Shanghai 200433, China}
\affiliation{Quantum Physics and Quantum Information Division, Beijing Computational Science Research Center, Beijing 100094, China}
\affiliation{Department of Applied Physics, Hong Kong Polytechnic University, Hung Hom,
Hong Kong, China}

\author{Da-Yu Jin}

\affiliation{Department of Physics, Fudan University, Shanghai 200433, China}

\author{Yueyin Qiu}

\affiliation{Department of Physics, Fudan University, Shanghai 200433, China}
\affiliation{Quantum Physics and Quantum Information Division, Beijing Computational Science Research Center, Beijing 100094, China}

\author{Chi-Hang Lam}

\altaffiliation{C.H.Lam@polyu.edu.hk}
\affiliation{Department of Applied Physics, Hong Kong Polytechnic University, Hung Hom,
Hong Kong, China}

\author{J. Q. You}

\altaffiliation{jqyou@csrc.ac.cn}
\affiliation{Quantum Physics and Quantum Information Division, Beijing Computational Science Research Center, Beijing 100094, China}
\affiliation{Synergetic Innovation Center of Quantum Information and Quantum Physics,
University of Science and Technology of China, Hefei, Anhui 230026, China}

\date{\today }

\begin{abstract}
  We study cross-Kerr (CK) effect on an optomechanical system driven by two-tone fields. We show that in the presence of the CK effect, a bistable behavior of the mean {photon number in the cavity} becomes more robust against the fluctuations of the frequency detuning between the cavity mode and the control field. The bistability can also be turned into a tri-stability within the experimentally accessible range of the system parameters. Also, we find that the symmetric profile of the optomechanically induced transparency is broken and the zero-absorption point is shifted in the presence of the CK effect. This shift can be used to measure the strength of the CK effect, and the asymmetric absorption profiles can be employed to engineer a high quality factor of the cavity.
\end{abstract}


\maketitle

\section{introduction}

Due to the potential applications to the study of a range of topics such as gravitational waves detection~\cite{VB} and tiny displacement measurement~\cite{DR}, optomechanical systems have received considerable theoretical and experimental interest in the past few years~\cite{MA}. In a typical setup, two mirrors (one is fixed and the other is movable) forming a cavity are subjected to a radiation pressure which causes the movable mirror to vibrate. This vibration in turn changes the length of the cavity (thus the frequency of the cavity modes) and gives rise to a nonlinear coupling between the cavity and the mechanical modes. To study the associated quantum effects, cooling of the nanomechanical resonator (NAMR) (i.e., the movable mirror) to the ground state is necessary. Various methods are proposed theoretically~\cite{huang1,wilson,liao} and recently successful cooling to ground state has been demonstrated experimentally~\cite{schliesser1,schliesser2,park}. However, experimental results~\cite{groblacher1,rocheleau,verhagen} demonstrate that single-photon optomechanical coupling cannot be operated in the strong coupling regime, indicating that non-trivial quantum effects cannot be observed at the single-photon level. Instead, a multi-photon strong coupling regime is accessible under a strong driving field on the cavity mode~\cite{teufel,groblacher2,akram}. Under this condition, quantum entanglement~\cite{paternostro,vitali,tian1}, quantum state transfer~\cite{wang,tian2,bochmann}, nonclassical states~\cite{mancini1,xu}, optomechanically induced transparency analogous to the electromagnetically induced transparency in atomic systems~\cite{agarwal,jing,ma}, and normal-mode splitting~\cite{dobrindt,huang2} have been investigated. However, realizing such a multi-photon strong coupling regime reduces the nonlinearity arising from the radiation pressure. The case with weak nonlinearity was discussed in studying photon blockade under both weak~\cite{rabl} and strong~\cite{nunnenkamp} driving fields. Also, it has been predicted that an optomechanical system driven by a strong field can give rise to an effective interaction Hamiltonian similar to that of a nonlinear Kerr medium interacting with photons~\cite{meystre,mancini2}, as investigated theoretically in~\cite{gong,aldana}. Therefore, {the hybridization of a Kerr medium and an optomechanical system driven by a strong field} can both enhance the optomechanical coupling and prevent the suppression of the nonlinearity. It has been proved that the Kerr nonlinearity in optomechanical systems can exhibit normal-mode splitting, reduce the photon number fluctuation, and provide a coherently controlled dynamics for the mechanical modes~\cite{kumar}. {Nevertheless, an external Kerr medium introduced to the system may also give rise to significant noise~\cite{lu1}.

Among various other suggested methods for attaining a strong coupling~\cite{lu2,heikkila,johansson,chesi,ludwig}, a promising approach~\cite{heikkila} involves an electromechanical system coupled to a two-level system such as a superconducting charge qubit~(see, e.g.,~\cite{you,xiang}), which has recently been demonstrated experimentally~\cite{pirkkalainen}. In this hybrid system, the optomechanical coupling can be enhanced several orders of magnitude and a controllable cross-Kerr (CK) effect is also introduced. Furthermore, Ref.~\cite{khan} showed that this CK effect can give rise to a sideband shift as well as an optimal cooling or heating. {Very recently, Ref.~\cite{sarala} reported a stability analysis of an optomechanical system with the CK effect, in which an optical bistable behavior was investigated.}

Motivated by these developments, we study the impacts of the CK effect on the steady-state behavior of the mean photon number~\cite{sete} and the optomechanically induced transparency~\cite{agarwal,jing,ma} in an optomechanical system. {Here we consider the case in which the strength of the CK effect has an opposite sign compared with that in Ref.~\cite{sarala}.}  We show that in the presence of the CK effect, the mean phonon number in the NAMR no longer varies monotonically with the mean photon number in the cavity, and a tri-stability behavior may be observed for the mean photon number. We also find that the condition to observe the bistable behavior is less stringent and more robust against the fluctuations of the frequency detuning between the cavity mode and the control field. Concerning optical absorption, without the CK effect, the absorption vanishes when the frequency detuning is comparable to the frequency of the NAMR, and the absorption profile with respect to the frequency detuning is symmetric about the zero-absorption point. However, in the presence of the CK effect, the zero-absorption point is shifted, which can be employed to measure the strength of the CK effect in the optomechanical system. Moreover, the absorption profile becomes asymmetric, with the linewidth of the left peak becoming much narrowed and the right one broadened. This indicates that a higher quality factor of the cavity can be engineered via introducing the CK effect.

The paper is organized as follows. In Sec.~II, the theoretical model is introduced and the Hamiltonian is given. Then, we use quantum Langevin equations to derive steady-state solutions in Sec.~III. Also, the steady-state behavior of the mean photon number is analyzed and the results with and without the CK effect are compared. In Sec.~IV, the CK effect on the optomechanically induced transparency is discussed. Finally, a conclusion is given in Sec.~V.

\section{The optomechanical system with CK effect}

{\subsection{Model and Hamiltonian}

\begin{figure}
\centering
  \includegraphics[scale=0.6]{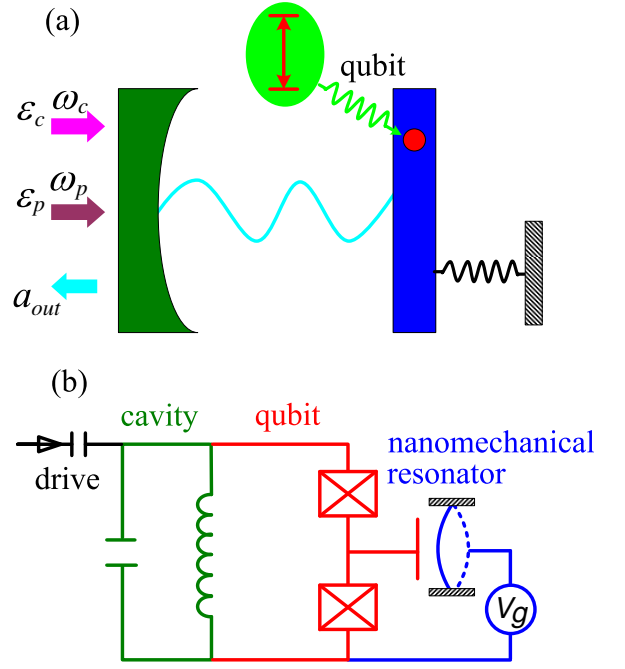}
  \caption{(Color online)~(a) Schematic diagram of the optomechanical system. One mirror of the cavity is fixed (green) and the other vibrates (blue). The assisting qubit (two-level system) (red) is used to induce the CK effect between the cavity and mechanical modes. (b) An equivalent quantum circuit~\cite{heikkila} with a superconducting charge qubit coupled to an on-chip cavity (e.g., a coplanar waveguide resonator) and a nanomechanical resonator. In (a), $\varepsilon_c$ and $\varepsilon_p$ are the amplitudes of the control and probe fields, respectively, while $a_{\rm out}$ is the amplitude of the output filed.}\label{opto}
\end{figure}
 A typical optomechanical system~\cite{MA} consists of a bare cavity and a NAMR, where the cavity mode is coupled to the NAMR mode via radiation pressure depending on both the photon number of the cavity and the displacement of the NAMR. The Hamiltonian of this optomechanical system can be written as (setting $\hbar=1$)~\cite{law}
\begin{equation}
H_{\rm opt}=\omega_a a^\dagger a+\omega_m b^\dag b-g_0 a^\dagger a (b+ b^\dag),\label{1}
\end{equation}
where $a^\dag~(a)$ is the creation (annihilation) operator of the cavity mode with frequency $\omega_a$, $b^\dag~(b)$ is the creation (annihilation) operator of the NAMR mode with frequency $\omega_m$, and $g_0$ is the coupling strength between the cavity and NAMR modes. Moreover, we consider the CK effect between the cavity and NAMR modes, corresponding to an interaction Hamiltonian of the form~\cite{heikkila}
\begin{equation}
H_{\rm ck}=-g_{\rm ck}a^{\dag}ab^{\dag}b.
\end{equation}
where $g_{\rm ck}$ is the CK coupling strength. This CK effect can be achieved via a two-level system [Fig.~\ref{opto}(a)] or a superconducting charge qubit [Fig.~\ref{opto}(b)] coupled to both the cavity and NAMR modes~\cite{heikkila,khan,pirkkalainen}. Also, the cavity is driven simultaneously by a control field with frequency $\omega_c$ and a probe field with frequency $\omega_p$ [see Fig.~\ref{opto}(a)]. The corresponding interaction Hamiltonian is
\begin{equation}
H_{\rm d}=(i\varepsilon_ce^{-i\omega_c t}+i\varepsilon_pe^{-i\omega_p t})a^{\dag}+{\rm H.c.},
\end{equation}
where $\varepsilon_{c(p)}=\sqrt{2\kappa \wp_{c(p)}/\hbar \omega_c}$ is the Rabi frequency between the cavity mode and the control (probe) field, with $\wp_c~(\wp_p)$ being the power of the control (probe) field and $\kappa$ the decay rate of the cavity mode induced by the thermal bath. Experimentally, $\varepsilon_p$ is usually chosen to be much smaller than $\varepsilon_c$.

The total Hamiltonian of the considered optomechanical system with CK effect can be written as $H_{\rm tot}=H_{\rm opt}+ H_{\rm ck}+H_{\rm d}$. Adopting the rotating frame with respect to the frequency of the control field, we apply a unitary transformation $S=\exp(-i\omega_c a^\dag a t)$ to the system. The total Hamiltonian is written as
\begin{eqnarray}
H\!&\!=\!&\!SH_{\rm tot}S^\dag+iS\partial_t S^\dag\nonumber\\
\!&\!=\!&\!\Delta_a a^{\dagger}a+\omega_m b^{\dagger}b-g_{0}a^{\dag}a (b^{\dagger}+b)\nonumber\\
&&\!-g_{\rm ck}a^{\dag}ab^{\dag}b+[(i\varepsilon_c+i\varepsilon_pe^{-i\Delta_p t})a^{\dag}+{\rm H.c.}],\label{RH}
\end{eqnarray}
where $\Delta_a=\omega_a-\omega_c$ is the frequency detuning of the cavity mode from the control field, and $\Delta_p=\omega_p-\omega_c$ is the frequency detuning between the probe and control fields.

\subsection{Quantum Langevin equations}

According to the Heisenberg-Langevin approach~\cite{walls}, the quantum dynamics of the considered optomechanical system can be described by the following quantum Langevin equations:
\begin{eqnarray}\label{QLE}
\dot a\!&\!=\!&\!-(i\Delta_a+\kappa)a+ig_0a(b+b^\dag)\notag\\
&&\!+ig_{\rm ck}ab^\dag b+\varepsilon_c+\varepsilon_pe^{-i\Delta_p t}+\sqrt{2\kappa}a_{\rm in},\\
\dot b\!&\!=\!&\!-(i\omega_m+\gamma)b+ig_0a^\dag a+ig_{\rm ck}a^\dag a b+\sqrt{2\gamma}b_{\rm in},\notag
\end{eqnarray}
where $\kappa$ ($\gamma$) is the decay rate of the cavity (NAMR) mode, and $a_{\rm in}$ ($b_{\rm in}$) is the input noise operator acting on the cavity (NAMR) mode, each of which has a zero mean value, i.e., $\langle a_{\rm in}\rangle=\langle b_{\rm in}\rangle=0$. Under Markov approximation, two-time correlation functions of these input noise operators are given by
\begin{eqnarray}
\langle a_{\rm in}(t) a_{\rm in}^\dag(t^\prime)\rangle\!&\!=\!&\!\delta(t-t^\prime),\notag\\
\langle b_{\rm in}^\dag(t) b_{\rm in}(t^\prime)\rangle\!&\!=\!&\!n_{\rm th}\delta (t-t^\prime),\\
\langle b_{\rm in}(t) b_{\rm in}^\dag(t^\prime)\rangle\!&\!=\!&\!(n_{\rm th}+1)\delta (t-t^\prime),\notag
\end{eqnarray}
where $n_{\rm th}=(e^{\hbar\omega_m/k_B T}-1)^{-1}$, with $k_B$ being the Boltzmann constant and $T$ the bath temperature, is the average phonon number in the thermal bath coupled to the NAMR. Here we write each of the system operators $a$ and $b$ as a sum of the steady-state value and the fluctuation via a factorization assumption, i.e., $a=\langle a\rangle+\delta a$, and $b=\langle b\rangle+\delta b$. It follows from Eq.~(\ref{QLE}) that the steady-state values of $a$ and $b$ satisfy
\begin{eqnarray}\label{sss}
\langle\dot a\rangle\!&\!=\!&\!-(i\Delta_a+\kappa)\langle a\rangle+ig_0\langle a\rangle(\langle b\rangle+\langle b^\dag\rangle)\notag\\
&&\!+ig_{\rm ck}\langle a\rangle\langle b^\dag\rangle\langle b\rangle+\varepsilon_c+\varepsilon_pe^{-i\Delta_p t},\\
\langle\dot b\rangle\!&\!=\!&\!-(i\omega_m+\gamma)\langle b\rangle+ig_0\langle a^\dag\rangle\langle a\rangle+ig_{\rm ck}\langle a^\dag\rangle\langle a\rangle\langle b\rangle\notag.
\end{eqnarray}
This mean-field approximation applies when the coupling between the cavity and NAMR modes is weak. In the weak driving regime ($\varepsilon_p\ll\varepsilon_c$), the nonlinear interaction between the cavity and NAMR modes affects the response of the system to the probe field. The solution to Eq.~(\ref{sss}) can be written as
\begin{eqnarray}
\langle a\rangle\!&\!=\!&\!A_0+A_+e^{-i\Delta_p t}+A_-e^{i\Delta_p t},\notag\\
\langle b\rangle\!&\!=\!&\!B_0+B_+e^{-i\Delta_p t}+B_-e^{i\Delta_p t},\label{assumption}
\end{eqnarray}
where $A_0$ ($B_0$) is the steady-state solution of the system operator $a$ ($b$) in the absence of the probe field ($\varepsilon_p=0$). For a weak probe field, $A_\pm$ ($B_\pm$) are much smaller than $A_0$ ($B_0$), but can have the same order of magnitude as the amplitude of the probe field. When the first-order terms are kept, the parameters in Eq.~(\ref{assumption}) can be analytically expressed as
\begin{eqnarray}\label{ssv}
A_0\!&\!=\!&\!\frac{\varepsilon_c}{\kappa+i\Delta},~~B_0=\frac{ig_0|A_0|^2}{\gamma+i\Omega_m},\notag\\
A_+\!&\!=\!&\!\frac{i(g^*B_++gB_-^*)A_0+\varepsilon_p}{\kappa+i\Delta_-},\notag\\
A_-\!&\!=\!&\!\frac{i(gB_+^*+g^*B_-)A_0+\varepsilon_p}{\kappa+i\Delta_+},\\
B_+\!&\!=\!&\!\frac{ig(A_0^*A_++A_0A_-^*)}{\gamma+i\omega_-},\notag\\
B_-\!&\!=\!&\!\frac{ig(A_0A_+^*+A_0^*A_-)}{\gamma+i\omega_+},\notag
\end{eqnarray}
where
\begin{equation}
\Delta=\Delta_a-g_{\rm ck}|B_0|^2-\frac{2g_0^2|A_0|^2\Omega_m}{\gamma^2+\Omega_m^2}
 \end{equation}
is the effective frequency detuning between the cavity mode and the control field in the optomechanical system with CK effect, which depends on the mean photon number $|A_0|^2$, the mean phonon number $|B_0|^2$, and the CK coupling strength $g_{\rm ck}$. Also, the radiation-pressure coupling strength $g_0$ is modified by the CK effect as $g=g_0+g_{\rm ck}B_0$. Other parameters in Eq.~(\ref{ssv}) are defined as $\Omega_m=\omega_m-g_{\rm ck}|A_0|^2$, $\Delta_\pm=\Delta \pm \Delta_p$, and $\omega_\pm=\Omega_m \pm \Delta_p$.

\begin{figure}
  \centering
  \includegraphics[scale=0.9]{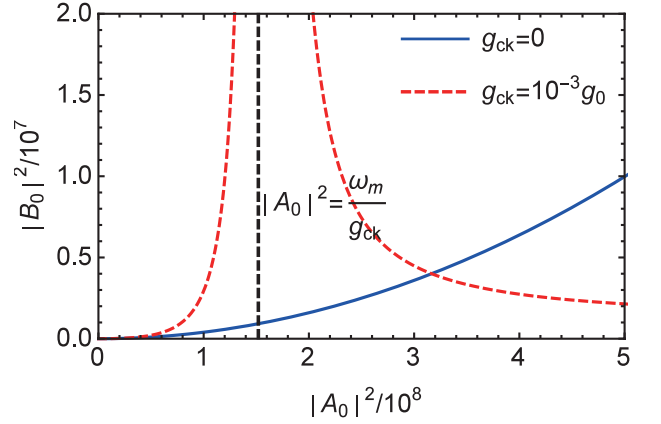}\\
  \caption{(Color online) Behavior of the mean phonon number $|B_0|^2$ as a function of the mean photon number $|A_0|^2$. The blue line corresponds to the usual optomechanical system without CK effect~($g_{\rm ck}=0$) and the red line corresponds to the optomechanical system in the presence of the CK effect~($g_{\rm ck}=10^{-3}g_0$). Other parameters are $\omega_a/2\pi=1.3$~GHz, $\omega_m/2\pi=6.3$~MHz, $g_0=250$~Hz, $\kappa/2\pi=0.1$~MHz, $\gamma=40$~Hz, and $\Delta_a=\omega_m$.}\label{pp}
  \end{figure}
In Fig.~\ref{pp}, we show the mean phonon number $|B_0|^2$ versus the mean photon number $|A_0|^2$ in the optomechanical systems with and without CK effect by using experimentally accessible parameters~\cite{rocheleau}:~$\omega_a/2\pi=1.3$~GHz, $\omega_m/2\pi=6.3$~MHz, $g_0=250$~Hz, $\kappa/2\pi=0.1$~MHz, $\gamma=40$~Hz, and $\Delta_a=\omega_m$. These parameters correspond to an optomechanical system in the resolved sideband regime where the frequency $\omega_m$ of the NAMR is larger than the decay rate $\kappa$ of the cavity. For the usual optomechanical system without CK effect, the mean phonon number increases monotonously with the mean photon number (see the blue curve in Fig.~\ref{pp}). However, the behavior of the mean phonon number differs drastically in the presence of the CK effect (see the red curve in Fig.~\ref{pp}). For $|A_0|^2<\omega_m/g_{\rm ck}$, the mean phonon number $|B_0|^2$ increases with the mean photon number $|A_0|^2$, but it then decreases with $|A_0|^2$ for $|A_0|^2>\omega_m/g_{\rm ck}$}.

\subsection{Steady-state behavior}

We now study the steady-state behavior of the mean photon number $|A_0|^2$ without the probe field ($\varepsilon_p=0$). In this case, using Eq.~(\ref{ssv}), it is straightforward  to show that $|A_0|^2$ satisfies
\begin{equation}
|A_0|^2\left[\kappa^2+\left(\Delta_a-g_0^2|A_0|^2\frac{g_{\rm ck}|A_0|^2+2\Omega_m}{\gamma^2+\Omega_m^2}\right)^2\right]=\varepsilon_c^2.\label{5}
\end{equation}
Due to the existence of the CK coupling $g_{\rm ck}$, this equation is more complicated than the case of the usual optomechanical system without CK effect~\cite{sete} . When the CK coupling vanishes, Eq.~(\ref{5}) is simplified to~\cite{sete}
\begin{equation}
|A_0|^2\left[\kappa^2+\left(\Delta_a-2g_0^2|A_0|^2\frac{\omega_m}{\gamma^2+\omega_m^2}\right)^2\right]=\varepsilon_c^2.\label{cubic}
\end{equation}
\begin{figure}
  \centering
  \includegraphics[scale=0.85]{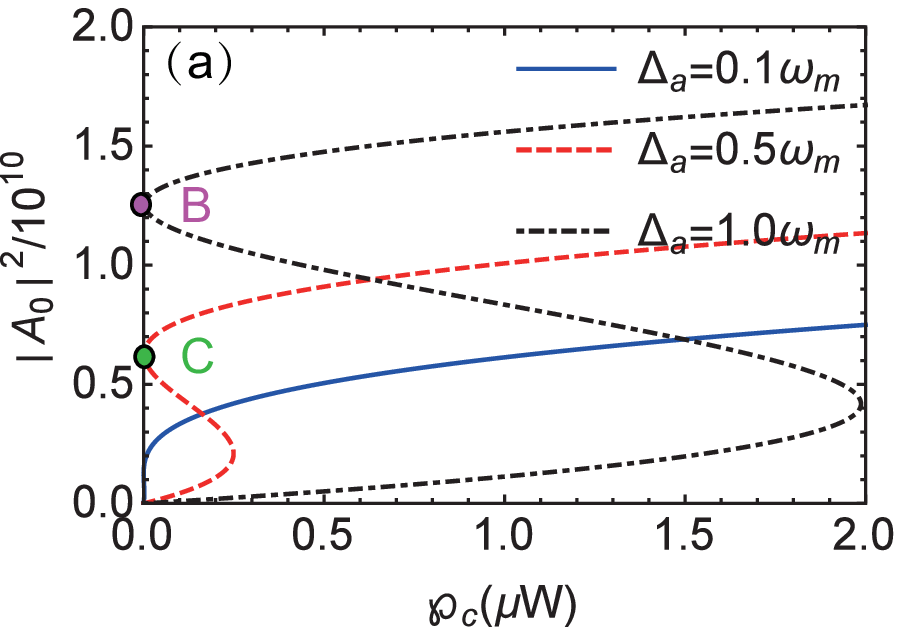}\\
  \includegraphics[scale=0.85]{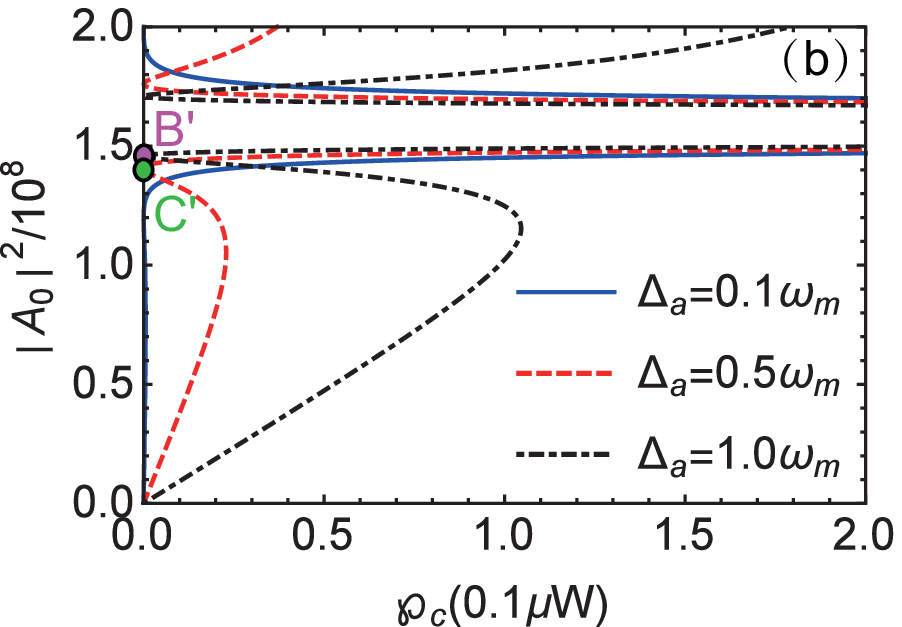}\\
  \caption{(Color online)~Mean photon number $|A_0|^2$ in the optomechanical systems with and without CK effect as a function of the control field $\wp_c$ for different values of $\Delta_a$. (a) The usual optomechanical system without CK effect ($g_{\rm ck}=0$). (b) The optomechanical system with CK effect ($g_{\rm ck}=0.001g_0$). Other parameters are the same as in Fig.~\ref{pp}.}\label{stabilityvsdetuning}
\end{figure}Note that Eq.~(\ref{cubic}) is a cubic equation in $|A_0|^2$ which can have up to three real roots. The leading nonlinear term originates from the  optomechanical coupling term $a^\dag a(b+b^\dag)$ which reduces to $|A_0|^2B_0$. In a certain parameter regime, the mean photon number thus exhibits a bistable behavior as shown by the black and red curves in Fig.~\ref{stabilityvsdetuning}(a).We now consider the CK effect that can be controlled by tuning the external gate voltage of a charge qubit in the optomechanical system~\cite{heikkila}. In this case,  Eq.~(\ref{5}) applies, which is a fifth-order equation and has up to five real roots. The higher nonlinearity is due to the CK coupling term $a^\dag ab^\dag b$ which leads to $|A_0|^2|B_0|^2$.  In additional to the bistable behavior, this can also gives rise to tri-stability in the optomechanical system. As shown by the black and red curves in Fig.~\ref{stabilityvsdetuning}(b), one can observe a tri-stable behavior of the mean photon number when appropriately tuning the power of the control field.

Below we determine the stability of the steady states of our system using the Routh-Hurwitz criterion. Note that $A_0$ ($B_0$) is the steady-state value of $\langle a\rangle$ ($\langle b\rangle$) in the absence of the probe field ($\varepsilon_p=0$). In this case, we can write $a$ and $b$ as $a\equiv\langle a\rangle+\delta a=A_0+\delta a$, and $b\equiv\langle b\rangle+\delta b=B_0+\delta b$. Substituting them into Eq.~(\ref{QLE}) and using Eq.~(\ref{sss}), we can obtain the dissipation-fluctuation equations of the motion as
\begin{eqnarray}\label{fluctuation}
\delta \dot a\!&\!=\!&\!-(i\Delta+\kappa)\delta a+i A_0(G^*\delta b+G\delta b^\dag)+\sqrt{2\kappa}a_{\rm in},\notag\\
\delta \dot b\!&\!=\!&\!-(i\Omega_m+\gamma)\delta b+i G(A_0^*\delta a+A_0\delta a^\dag)+\sqrt{2\gamma}b_{\rm in},\notag\\
\end{eqnarray}
where
\begin{equation}
G=g_0 \left(1+\frac{ig_{\rm ck}|A_0|^2}{\gamma+i\Omega_m}\right).
\end{equation}
In a compact matrix form, Eq.~(\ref{fluctuation}) can be recast as
\begin{align}
\delta {\bf \dot v}=&{\bf C}{\bf v}+\delta {\bf v_{\rm in}}
\end{align}
with the operator vectors ${\bf v}=\big(\delta a,~\delta a^\dag,~\delta b,~\delta b^\dag\big)^{\rm T}$, and $\delta {\bf v_{\rm in}}=\big(\sqrt{2\kappa} a_{\rm in},~\sqrt{2\kappa}a_{\rm in}^\dag,~\sqrt{2\gamma}b_{\rm in},~\sqrt{2\gamma} b_{\rm in}^\dag\big)^{\rm T}$, where $T$ denotes the transpose of a matrix. The matrix ${\bf C}$ is given by
\begin{equation}
{\bf C}=\left(
           \begin{array}{ccccc}
             -i\Delta-\kappa & 0 & iA_0G^* & iA_0G \\
             0 & i\Delta-\kappa& -iA_0^*G & -iA_0^*G^* \\
             iA_0^*G & iA_0G & -i\Omega_m-\gamma & 0 \\
             -iA_0G^* & -iA_0^*G^* & 0 & i\Omega_m-\gamma \\
           \end{array}
         \right).
\end{equation}
The characteristic equation $| {\bf C} - \lambda {\bf I} |  = 0$ can be reduced to $\lambda^4+C_3\lambda^3+C_2\lambda^2+C_1\lambda+C_0=0$, where the coefficients can be derived using straightforward but tedious algebra.
For the particular case with $\gamma+i\Omega_m\approx i\Omega_m$, their explicit expressions can be found in Ref.~\cite{sarala}. From the Routh-Hurwitz criterion~\cite{routh}, a solution is stable only if the real part of the corresponding eigenvalue $\lambda$ is negative, and the stability conditions can then be obtained as
\begin{eqnarray}
C_3\!&\!>\!&\!0,\nonumber\\
C_3C_2-C_1\!&\!>\!&\!0,\\
C_3C_2C_1-(C_1^2+C_3^2C_0)\!&\!>\!&\!0.\nonumber
\end{eqnarray}
Using our model parameters and considering $|A_0|^2>\omega_m/g_{\rm ck}$, we find that three positive roots of Eq.~(\ref{5}) correspond to the stable solutions, leading to a tri-stable behavior. The remaining two positive roots sandwiched in between correspond to the unstable solutions as expected.
In the Appendix, we have also verified using Descartes rule that at most five positive roots can exist for Eq.~(\ref{5}). We should also point out that a similar analysis in Ref.~\cite{sarala}
considering, in our notation, a negative value of $g_{\rm ck}$ leads to a bistable behavior.

Interestingly, the onset of bistability requires a weaker control field $\wp_c$ in the presence of the CK effect, so the condition of bistable behavior for the mean photon number in an optomechanical system with the CK effect is more relaxed. We now explain the effect of the frequency detuning on the bistable behavior with or without CK effect. From Figs.~\ref{stabilityvsdetuning}(a) and~\ref{stabilityvsdetuning}(b), a relatively large detuning $\Delta_a=\omega_a-\omega_c$ is needed to observe the bistable phenomenon. For instance, the bistable phenomenon does not occur at $\Delta_a=0.1\omega_m$ (see the blue curves in Fig.~\ref{stabilityvsdetuning}). However, the stable value of the mean photon number is more robust against the variation of the detuning value in the presence of the CK effect. For example, without CK effect, the upper stable point shifts significantly from B to C as $\Delta_a$ decreases~[Fig.~\ref{stabilityvsdetuning}(a)], but it shifts only a little from B$^\prime$ to C$^\prime$
in the presence of the CK effect [Fig.~\ref{stabilityvsdetuning}(b)]. Therefore, the CK effect in the optomechanical system can be used to suppress the variation of the bistable point with respect to the frequency detuning between the cavity mode and the control field.

\section{CK effect on the optomechanically induced transparency}

\begin{figure}
  \includegraphics[scale=0.85]{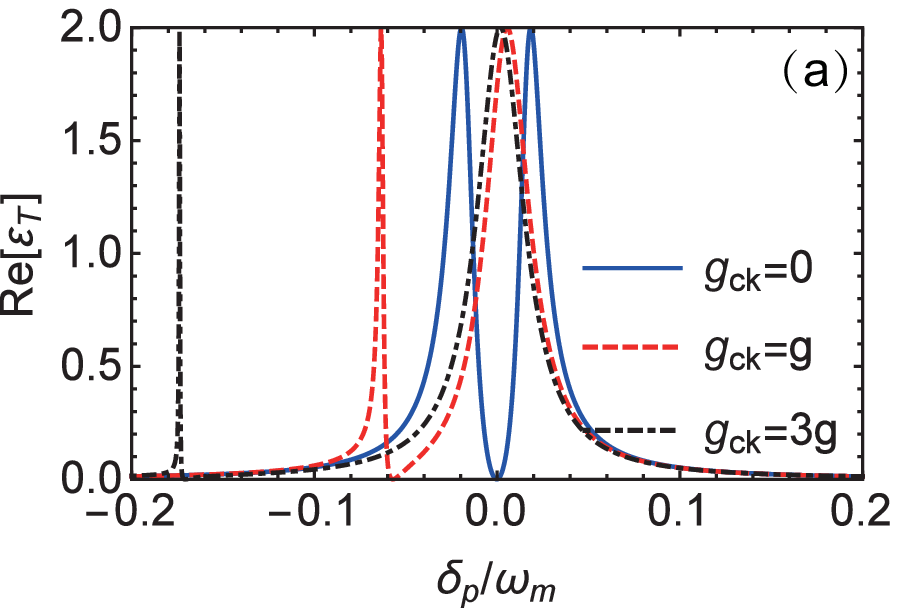}\\
  \includegraphics[scale=0.86]{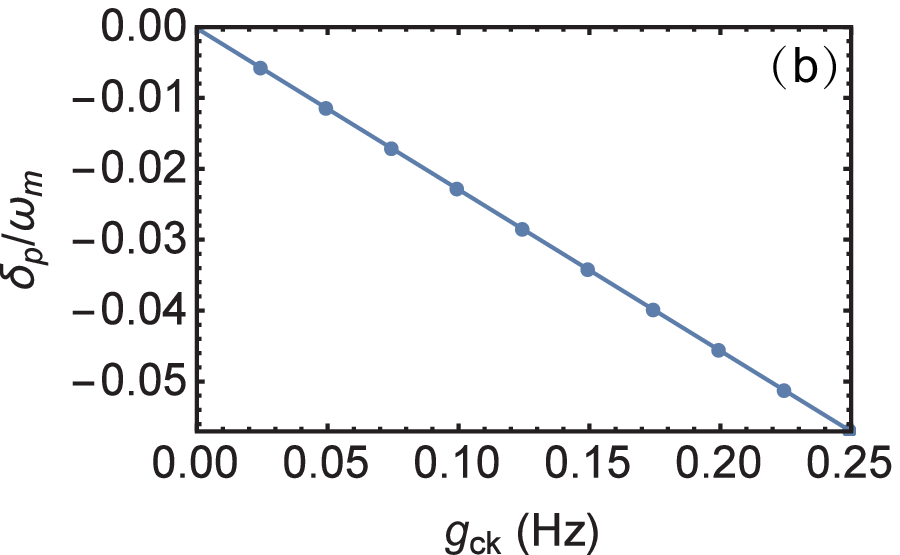}\\
  \caption{(Color online) (a) Absorption of the probe field (i.e., Re[$\varepsilon_T$]) as a function of the reduced detuning $\delta_p/\omega_m$ for different values of $g_{\rm ck}$, where the control field is fixed at $\wp_c=9.6$ nW and $g=10^{-3}g_0$. (b) Probing CK effect based on optomechanically induced transparency phenomenon. Other parameters are the same as in Fig.~\ref{pp}.}\label{omitvscoupling}
\end{figure}
  Below we investigate the transmission of a probe field through an optomechanical system in the presence of the CK effect. Using input-output theory~\cite{walls}, we obtain
 \begin{equation}
 \langle a_{\rm out}\rangle+(\varepsilon_c+\varepsilon_pe^{-i\Delta_p t})/\sqrt{2\kappa}=\sqrt{2\kappa}\langle a\rangle. \label{theory}
 \end{equation}
 Also, the output of the cavity field can be expressed in a similar form to Eq.~(\ref{assumption}), i.e.,
 \begin{equation}
 \langle a_{\rm out}\rangle=A_{\rm out}+A_{\rm out}^+e^{-i\Delta_p t}+A_{\rm out}^-e^{i\Delta_p t}.\label{ass2}
 \end{equation}
 By comparing Eq.~(\ref{ass2}) with Eq.~(\ref{theory}), we have
 \begin{eqnarray}\label{AOUT}
 A_{\rm out}\!&\!=\!&\!\sqrt{2\kappa}A_0-\varepsilon_c/\sqrt{2\kappa},\notag\\
  A_{\rm out}^+\!&\!=\!&\!\sqrt{2\kappa}A_+-\varepsilon_p/\sqrt{2\kappa},\\
  A_{\rm out}^-\!&\!=\!&\!\sqrt{2\kappa}A_-,\notag
\end{eqnarray}
 where $A_{\rm out}$ is the output {amplitude at the control field  frequency $\omega_c$}, $A_{\rm out}^+$ is the output amplitude at the Stokes frequency (i.e. probe frequency) $\omega_p$, and $A_{\rm out}^-$ is the output amplitude at the anti-Stokes frequency $2\omega_c-\omega_p$. Now we define a reduced output field which responses to the probe field at frequency $\omega_p$ as
 \begin{equation}
 \varepsilon_T\equiv\frac{\sqrt{2\kappa}A_{\rm out}^+}{\varepsilon_p}+1=\frac{2\kappa A_+}{\varepsilon_p},\label{output}
 \end{equation}
 where
 \begin{equation}
 A_+=\frac{s/(\kappa+i\Delta_-)+i|g|^2|A_0|^2(\omega_++\omega_-)}{s-|g|^2|A_0|^2(\omega_++\omega_-)(\Delta_++\Delta_-)}\varepsilon_p,
 \end{equation}
as given by Eq.~(\ref{ssv}), and $s=(\gamma+i\omega_-)(\gamma-i\omega_+)(\kappa+i\Delta_-)(\kappa-i\Delta_+)$. The real and imaginary parts of $\varepsilon_T$ describe  the absorption and the dispersion of the probe field, respectively, which can be  measured via, e.g., homodyne detections~\cite{walls}.
Figure~\ref{omitvscoupling}(a) displays the real part of the output field $\varepsilon_T$  defined by Eq.~(\ref{output}) as a function of the reduced detuning $\delta_p/\omega_m$, where $\delta_p=\Delta_p-\omega_m$. For the usual optomechanical system without CK effect ($g_{\rm ck}=0$), the absorption Re[$\varepsilon_T$] vanishes at $\Delta_p=\omega_m$ and the absorption is symmetric about $\delta_p=0$ [see the blue curve in Fig.~\ref{omitvscoupling}(a)]. This optomechanically induced transparency phenomenon~\cite{agarwal,jing,ma} is similar to the electromagnetically induced transparency in atomic systems. When the CK effect is introduced, the position of the zero-absorption point is shifted and the absorption curve becomes asymmetric. When increasing the strength $g_{\rm ck}$ of the CK effect, the absorption peak on the left becomes much narrowed, and the one on the right is broadened~[see the red and black curves in Fig.~\ref{omitvscoupling}(a)]. This result suggests that the CK effect can be employed to improve the quality factor of the cavity when the reduced detuning $\delta_p/\omega_m$ is tuned to be around the left peak. In addition, the separation between the two peaks is widened as $g_{\rm ck}$ increases. In Fig.~\ref{omitvscoupling}(b), we show the position of the zero-absorption point against the strength of the CK effect. The simple dependence shown in Fig.~\ref{omitvscoupling}(b) indicates that the strength of the CK effect can be easily probed by measuring the zero-absorption point.

\section{conclusion}

{In conclusion, we have studied the impacts of the CK effect on the steady-state behavior of the mean phonon number and optical transparency in an optomechanical system. In this nonlinear system, the mean phonon number does not vary monotonically with the mean photon number, and a tri-stability for the mean photon number under realistic conditions may be observed. We also find that the bistable behavior of the photon number becomes more robust against the variation of the system parameters compared to the case without the CK effect. Concerning the optomechanically induced transparency phenomenon, we have predicted an asymmetric absorption profile and a shift of the zero-absorption point when the detuning between the cavity mode and the control field is comparable to the frequency of the NAMR. This asymmetry can be employed to improve the quality factor of the cavity, and the shift of the zero-absorption point can be used to probe the strength $g_{\rm ck}$ of the CK effect.

\section*{ACKNOWLEDGMENTS}

This work is supported by the NSAF No.~U1330201, the National Natural Science Foundation of China No.~91421102, the National Basic Research Program of China No.~2014CB921401, and the Hong Kong GRF~No.~501213.
\vskip 0.5 cm

\setcounter{equation}{0}
\renewcommand{\theequation}{A\arabic{equation}}
\appendix{}

\section*{APPENDIX}

In this Appendix, we provide a direct and efficient estimation on how many positive solutions exist for Eq.~(\ref{5}) according to Descartes rule. Because
$\Omega_m=\omega_m-g_{\rm ck}|A_0|^2$ and $\gamma^2+\Omega_m^2\neq0$, Eq.~(\ref{5}) can be recast as
\begin{equation}\label{x}
a_5x^5+a_4x^4+a_3x^3+a_2x^2+a_1x+a_0=0,
\end{equation}
where we define $x\equiv|A_0|^2$, and the coefficients are
\begin{eqnarray}\label{coeff}
a_0\!&\!=\!&\!-\varepsilon_c^2 (\gamma^2+\omega_m^2 )^2,  \notag\\
a_1\!&\!=\!&\!(\gamma^2+\omega_m^2 )[4\varepsilon_c^2 g_{\rm ck} \omega_m+(\Delta_a^2+\kappa^2 )(\gamma^2+\omega_m^2 )], \notag\\
a_2\!&\!=\!&\!-2\{2[\Delta_a (g_0^2+\Delta_a g_{\rm ck} )+g_{\rm ck} \kappa^2 ] \omega_m (\gamma^2+\omega_m^2 ) \notag\\
&&\!+\varepsilon_c^2 g_{\rm ck}^2 (\gamma^2+3\omega_m^2 )\},\\
a_3\!&\!=\!&\!2g_{\rm ck} [\Delta_a (g_0^2+\Delta_a g_{\rm ck} )+g_{\rm ck} \kappa^2 ] \gamma^2+4\varepsilon_c^2 g_{\rm ck}^3 \omega_m  \notag\\
&&\!+2[2g_0^4+5\Delta_a g_0^2 g_{\rm ck}+3g_{\rm ck}^2 (\Delta_a^2+\kappa^2 )] \omega_m^2, \notag\\
a_4\!&\!=\!&\!-g_{\rm ck} \{\varepsilon_c^2 g_{\rm ck}^3+4[(g_0^2+\Delta_a g_{\rm ck} )^2+g_{\rm ck}^2 \kappa^2] \omega_m \}, \notag\\
a_5\!&\!=\!&\!g_{\rm ck}^2 (g_0^2+\Delta_a g_{\rm ck} )^2+g_{\rm ck}^4 \kappa^2. \notag
\end{eqnarray}
Because all parameters $g_0$, $g_{\rm ck}$, $\omega_m$, $\Delta_a$, $\varepsilon_c$, $\kappa$, and $\gamma$ in Eq.~(\ref{coeff}) are positive, we have $a_0<0$, $a_1>0$, $a_2<0$, $a_3>0$, $a_4<0$, and $a_5>0$, corresponding to the following unique sign sequence:
\begin{equation}
\text{sign}(a_5)\ldots \text{sign}(a_0)=+-+-+-.
\end{equation}
According to Descartes rule, Eq.~(\ref{x}), namely, Eq.~(\ref{5}) in the main text, has at most five positive roots, which are consistent with our results [see the red and black curves in Fig.~\ref{stabilityvsdetuning}(b)].

We should point out that a different case with the definition $\omega_m^e=\Omega_m=\omega_m+g_{\rm ck}|A_0|^2=\omega_m+g_{\rm ck} x$, where $g_{\rm ck}>0$, was studied in Ref.~\cite{sarala}, which is equivalent to a negative $g_{\rm ck}$ in our notation. This gives a bistable behavior as shown in Ref.~\cite{sarala}, and an application of Descartes rule to it gives at most three positive roots instead as expected.

\end{document}